 \documentstyle[12pt]{article}
 
\begin{document}

 \rightline{EFI 95-58}

 \begin{center} \large {\bf Discrete Scaling in Stock Markets Before Crashes}

 \medskip\normalsize
                 James A. Feigenbaum and Peter G.O. Freund\\

             {\em Enrico Fermi Institute and Department of Physics\\
             The University of Chicago, Chicago, IL 60637, USA}\\

 \end{center}
 \medskip

\noindent{\bf Abstract}: We propose a picture of stock market crashes as
critical points in a hierachical system with discrete scaling. The critical
exponent is then complex, leading to log-periodic fluctuations in stock
market indexes. We present ``experimental'' evidence in favor of this
prediction. This picture is in the spirit of the known earthquake-stock
market analogy and of recent work on log-periodic fluctuations associated
with earthquakes.

\bigskip
\bigskip
\bigskip
\bigskip

The study of earthquakes as critical points has been of interest for some time
now \cite{CP, ALL, STS, TS, N}. At a critical point one expects a scaling
regime to set in.
Recently it has been suggested \cite{JS,SSS} that the underlying scale
invariance is discrete, as expected for a hierarchical system.
Then the critical exponent is complex and the
scaling law near the critical point is ``decorated'' by log-periodic
corrections ($\Re\tau^{\alpha +i\omega}=\tau^{\alpha}\cos( \omega \log\tau)$)
Evidence for such log-periodic fluctuations was found \cite{JS}
in measurements of the concentration
of $Cl^{-}$ and $SO_{4}^{--}$ ions in mineral water collected over the 20
months
immediately preceeding the 1995 Kobe earthquake at a source close to its
epicenter. Similar evidence was also found \cite{SSS} in the cumulative
Benioff strain in connection with the 1989 Loma Prieta earthquake. It was
proposed \cite{JS} that monitoring log-periodic
fluctuations may ultimately prove useful in earthquake prediction.

Such fluctuations seem generic in hierarchically organized rupture processes.
In the spirit of an earthquake-stock market analogy, this has led
us to consider the possibility that log-periodic fluctuations may appear
in stock market indices over a period preceding a crash. The stock market index
(S\&P 500, Dow-Jones, NIKKEI, ...) is to play here the same role as the
$Cl^{-}$ ion concentration played in the the analysis of the Kobe earthquake.
Fortunately such indices are
closely monitored and good data are plentiful. The scaling variable is again
time $t$. Call $c(t)$ the index as a function of time. Truncating
at the first harmonic of a general log-periodic correction, we can then
write for $c(t)$ the same formula as that given in \cite{JS} for the ion
concentration
$$
c(t)= A+ B(t_{c}-t)^{\alpha} \left[ 1+ C \cos[\omega
\log(t_{c}-t)+\varphi]\right]. ~ ~ ~
\eqno(1)
$$

As a first test of this idea let us consider the crash which occured in New
York on October 19, 1987. As the relevant index let us use the S\&P 500,
which dropped
by more than 20\% that day. In figure 1a we present a fit of the 1986-1987
weekly S\&P 500 using  Eq. (1) . One can see clearly two full periods
of the log-periodic oscillation and some more oscillatory behavior close to
the time of the crash. We only fit data up to three weeks before the crash,
where
the fit starts very fast oscillations. A reasonable fit is obtained this way
with parameters given in Table 1 (where we omitted the parameters $A$,
$B$ and $\varphi$, which depend
on the arbitrary normalization of the index or on the time scale).

Error bars of $\pm 10$ were
assigned to each data point for purposes of calculating $\chi^2$. To a certain
extent this is arbitrary, but it also reflects the possibility of higher
harmonics neglected in our fit and of noise. This error assignment will be
used in all other fits except the next one. Concerning the values of $t_c$
for this and our other fits as well, the actual crash dates have been used as
input.
Given that here we used weekly, and in all other fits monthly, data, the small
discrepancies between the crash dates in Table 1 (given there in the
yy/mm/dd format) and the real crash dates
are devoid of significance.

In figure 1b we fit a NY crash (defined here as a drop in the index by
more than 10\%) in 1962, using monthly S\&P 500 data. This time the error
bars used to calculate $\chi^2$ were set at $\pm 2.5$, since the S\&P 500
was considerably lower in the Fifties than in the Eighties.
In figure 1c the 1929 NY crash (using monthly Dow-Jones data) is fit.
The parameters for all these fits are given in table 1.
We also considered the 1990 Tokyo crash using {\em scanned} NIKKEI data
and found similar log-periodic behavior, but we intend to refine this with
tabulated data.

There is evidence for log-periodicity in these fits.
While log-periodicity is indicative of an impending crash,
one can easily think of crashes not associated with any log-periodic behavior,
for instance those caused by sudden unexpected world events.
We should add that we defined a crash as a change by
10\% or more of an index over a short time interval (e.g 1 day in 1987).
In all cases the change is negative (a crash) rather than positive (an
upsurge).

Notice that all these fits range over time intervals of 2-8 years
before the crash. By contrast, let us attempt to fit the S\&P 500 for
the time interval 1991-1994
during (and immediately after) which no crash occured. This is
done in figure 1d and its
parameters can be found in table 1. The parameter $C$ which
measures the relative importance of the log-periodic fluctuations is now two
orders of magnitude smaller than in all previous fits
and that there are therefore no detectable oscillations.
The critical time $t_c$ itself is
in the {\em past}. Restricting to a shorter calm period, say 1992-1993,
no significant oscillations are observed either.
This further supports the discrete scaling picture advocated here.

Let us now return to the 1987 crash. We have fitted Eq. (1) to the
1986-1987 interval leading up to this crash. One might discern further
periods of the log-periodic fluctuations over a longer time period.
In fact if we select the interval 1980-1987, then six oscillation periods
come into view. Can one fit these to equation (1)? The answer is yes and
the corresponding best fit is given in Fig. 1e and the parameters
again in table 1. There is a problem now, for the frequency $\omega$ is now
quite different from that obtained from the 1986-1987 fit. One might think
that relaxing the
assumed constancy of the background parameter $A$ might alleviate this
problem. Yet
assuming A to be a quadratic polynomial in time (at the expense of two added
parameters) has no significant effect.
The eight-year
fit involves a higher frequency $\omega$, which makes it overoscillate in the
overlap region with the two-year fit, so that over the final years 1986 and
1987 a new complex critical exponent takes over.

The next phenomenological question concerns the accuracy with which the time
$t_c$ of the crash can be predicted from the monitoring of log-periodic
fluctuations. This is an interesting question indeed and
through much more detailed statistical study one could settle it
for past crashes. The authors of ref. \cite{JS} have expressed optimism
concerning the corresponding problem for earthquakes, namely predicting
earthquakes on the basis of monitoring
log-periodic fluctuations at the ``right'' sites. But seismic activity
is a natural phenomenon impervious to human monitoring. By contrast, if in
the future
large groups of investors, who believe they have observed a pattern of
log-periodic fluctuations in a stock market index, proceed
on that basis to
predict a crash time, they may find such a prediction both unprofitable and
``counterproductive''.

As a rule, discrete scaling is connected with hierarchical models
\cite{SSS}. It is thus natural to invoke a hierarchical structure
to account for the discrete scaling connected with log-periodic
fluctuations in stock market indexes. A clear hierarchical structure
is present among investors which range from the individual small investor
to the  largest mutual funds. At the other end, the stocks themselves
arrange themselves in subsectors, sectors, industries, ...
A fiber bundle-like model \cite{N} exploiting these hierarchies can be
envisaged and we hope to return
to this subject elsewhere. Here we prefer to keep the discussion
``phenomenological'' and content ourselves with the above presentation of
evidence for log-periodic fluctuations in stock market indexes .

While this work was being completed, we learned that D. Sornette and
coworkers have also considered this problem with similar results.

\medskip

{\bf Acknowledgements}\\
We wish to thank Caroline L. Freund, Jerry F. Gould,
Michael Isenberg, Leo Kadanoff, Sidney
Nagel, Tom Witten and Cosmas Zachos for valuable discussions and Hubert
Saleur for sending us preprints of references \cite{JS} and \cite{SSS}.
We are grateful to Aaron K. Grant of UCLA and to Greg Howell of Investment
Research, Chicago for their generous help with the fits and for very useful
discussions. We thank David Fialkowski of CRSP for supplying us
with S\&P 500 data. This work was supported in part by NSF grants.
While most of this work was done, one of us (P.G.O.F) was visiting the HEP
Division at the Argonne National Laboratory. He wishes to acknowledge
the generous hospitality extended to him there.
\newpage

        \begin{center} {\bf Table 1}

        \end{center}

\begin{tabular}{|l|l|l|l|l|l|l|r|}   \hline
years/index & figure   & $C$    & $\alpha$ & $\omega$ & $t_c$ &
$\chi^2$/degrees freedom \\ \hline
86-87/S\&P 500      &  1a     & -0.035 &    0.2   &   8.06   & 87/10/19 &
45.41/76    \\ \hline
53-62/S\&P 500   &  1b     &  0.19 &    0.68  &    7.41  & 62/01/04  & 78.17/88
 \\ \hline
20-29/DJ      &  1c     & -0.014 &   0.14   &    8.73  & 29/10/22   & 101/103
\\ \hline
91-94/S\&P 500      &  1d     &  0.00058 &  0.57  &   12.01  & 90/05/07 &
60.79/41  \\ \hline
80-87/S\&P 500      &  1e     &  -0.036 &    0.2   &   12.94   & 87/10/15   &
99.47/84  \\ \hline
\end{tabular}

\newpage

\begin{center}  {\bf Figure Caption}

\end{center}

{\bf Figure 1}.
Fits with Eq. (1) of the: a) 1986-1987 S\&P 500 ;
b) 1953-1962 S\&P 500; c) 1920-1929 Dow-Jones;
d) 1990-1994 S\&P 500; e) 1980-1987 S\&P 500. The values of the parameters
and the $\chi^2$ for each of these fits are given in Table 1.

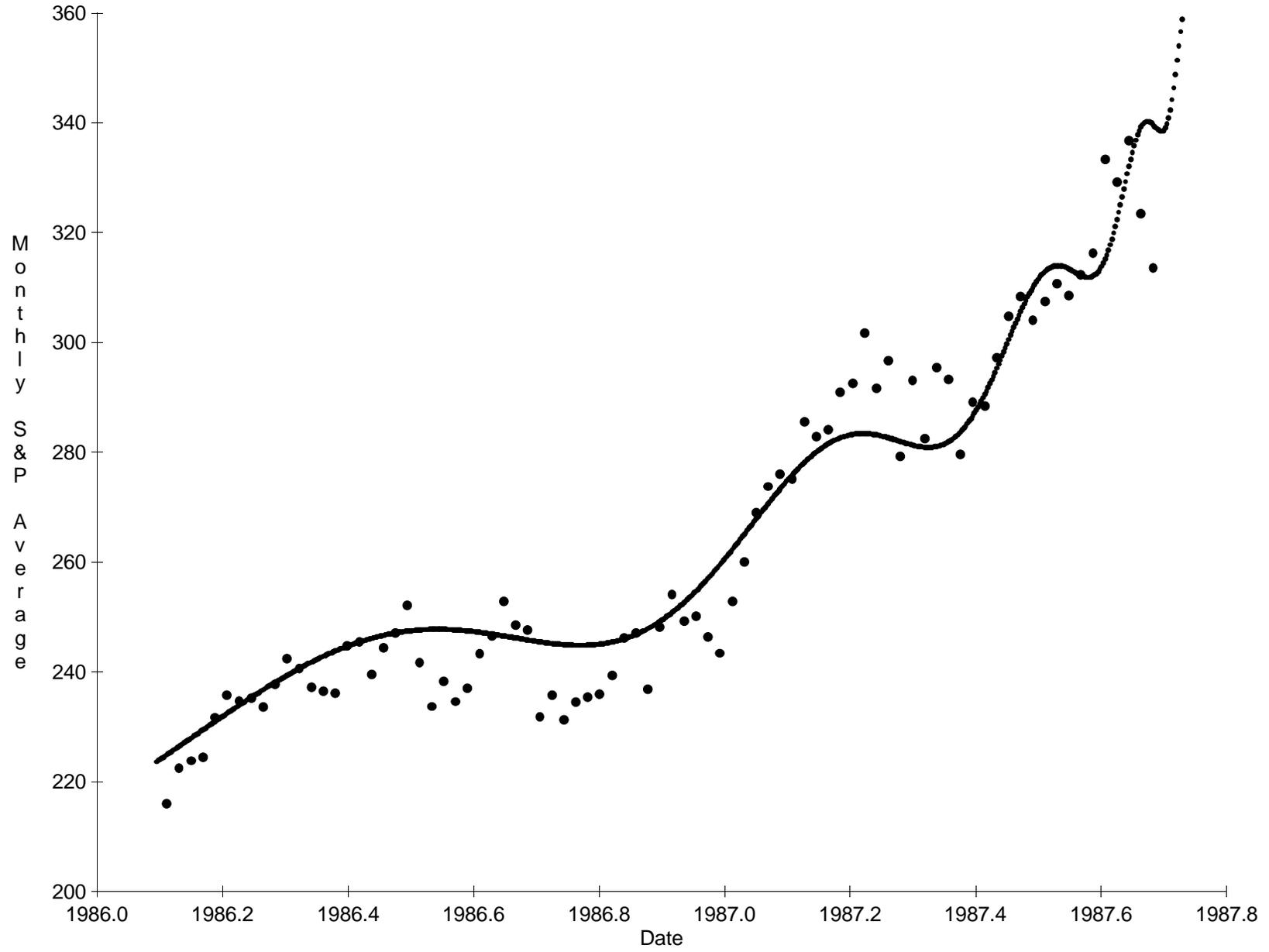
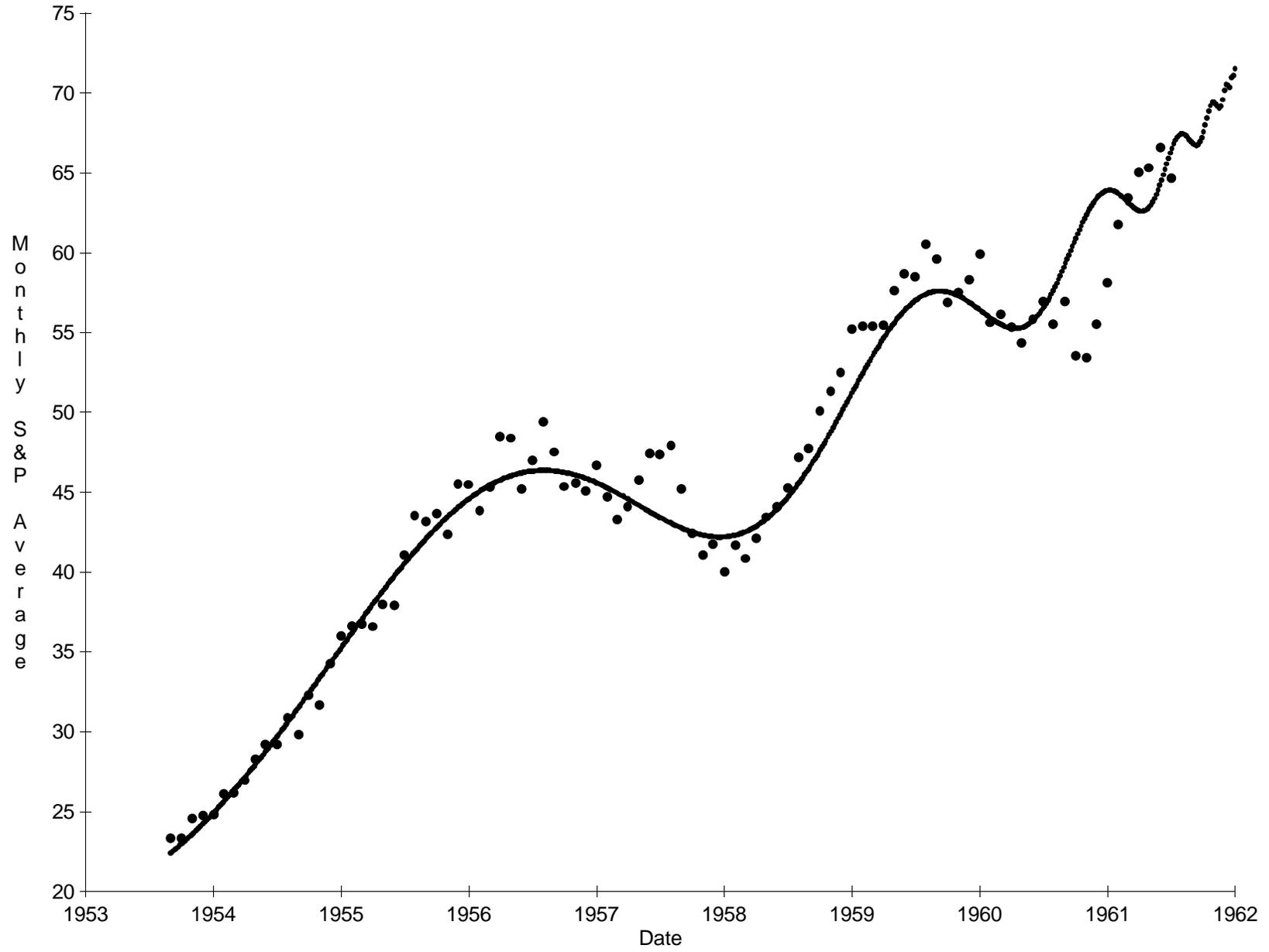
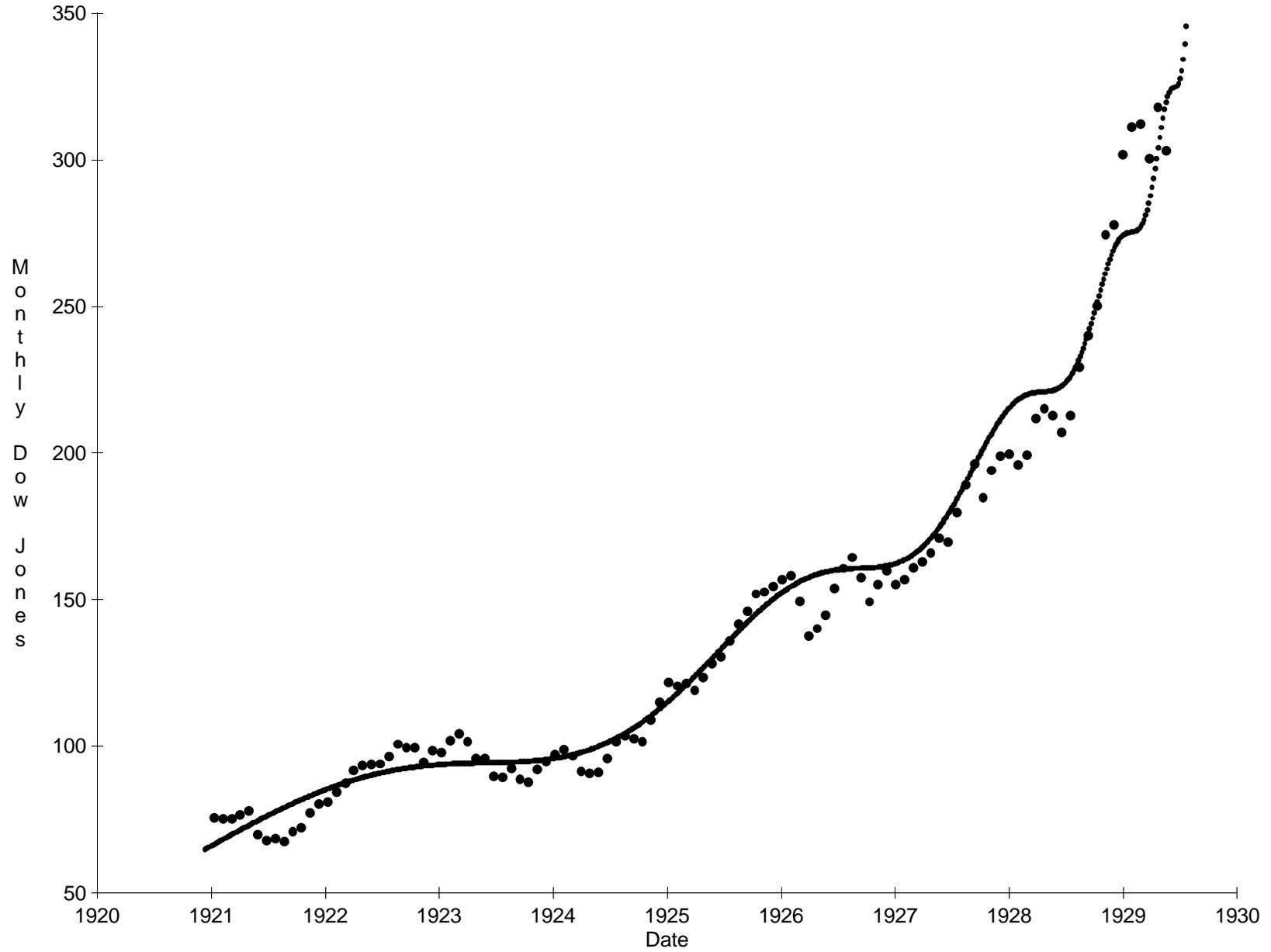
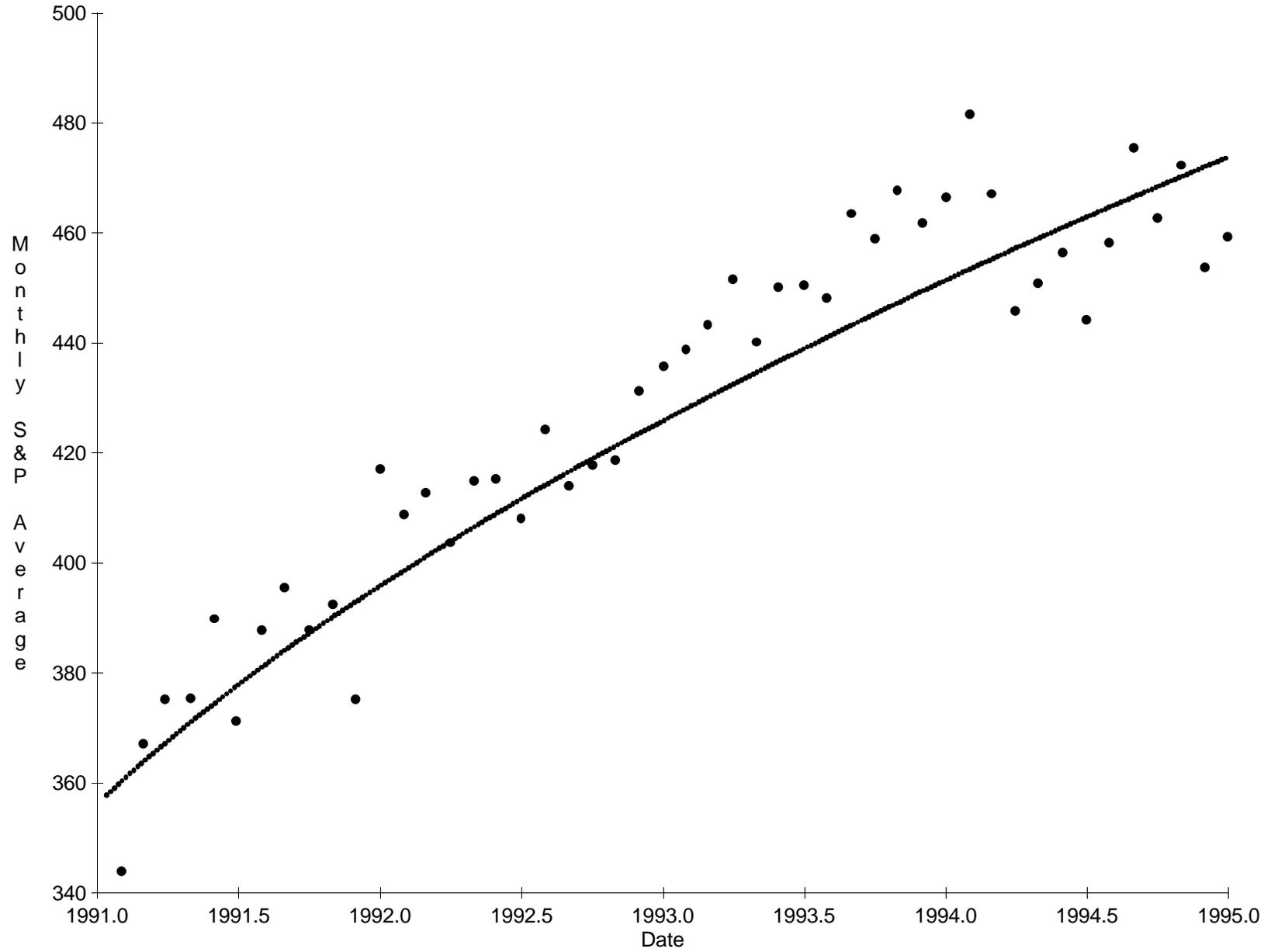
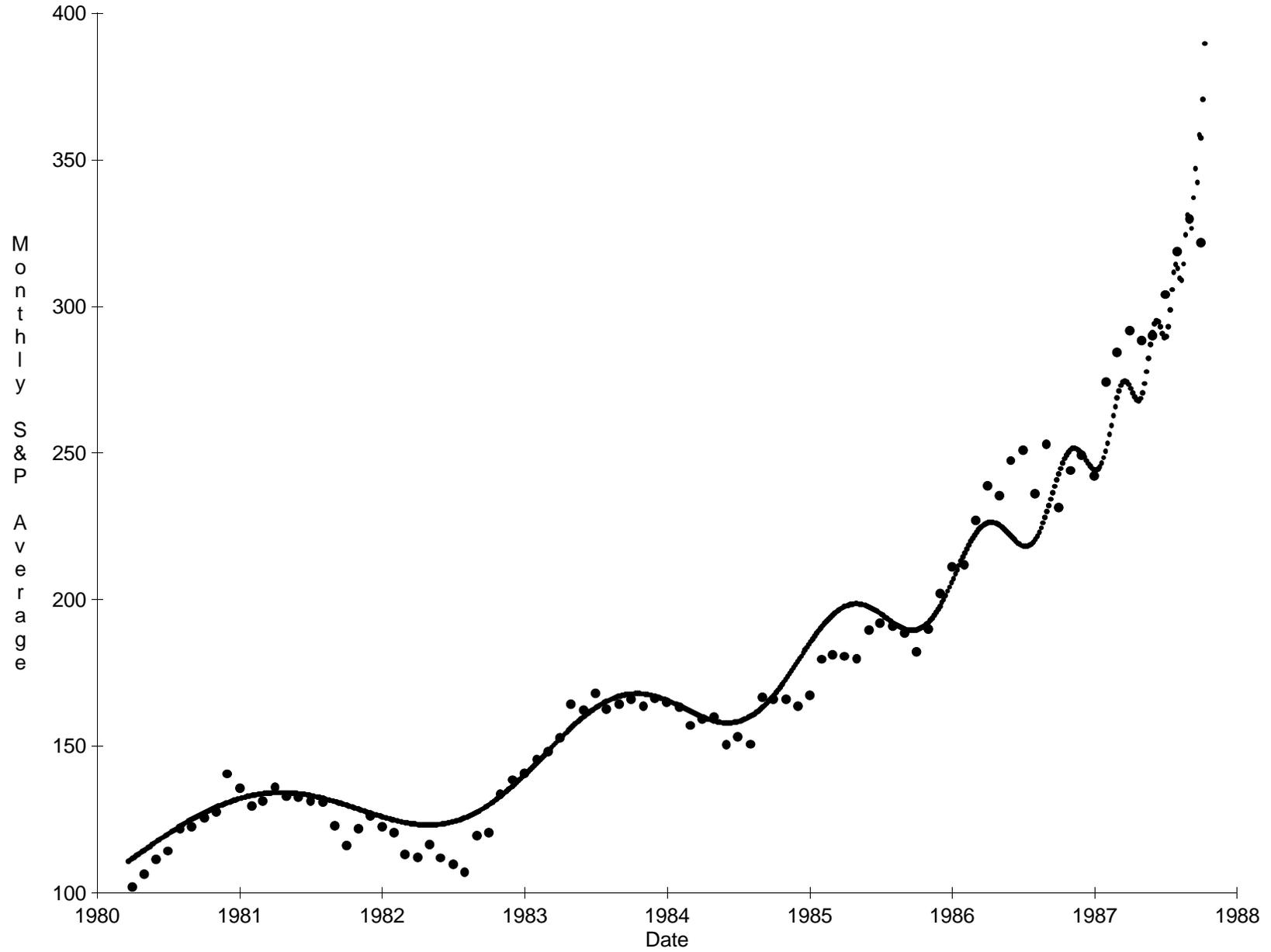
\end{document}